\documentstyle[prl,floats,aps,twocolumn,epsf,graphicx]{revtex}
\begin{document}
\twocolumn[\hsize\textwidth\columnwidth\hsize\csname
@twocolumnfalse\endcsname

\title{Intermediate Asymptotics of the Kerr Quasinormal Spectrum}
\author{Shahar Hod$^1$ and Uri Keshet$^2$}
\address{$^1$The Racah Institute of Physics, The Hebrew University, Jerusalem 91
904, Israel}
\address{}
\address{$^2$Physics Faculty, Weizmann Institute, Rehovot 76100, Israel}
\date{\today}
\maketitle

\begin{abstract}

\ \ \ We study {\it analytically} the quasinormal mode spectrum of near-extremal 
({\it rotating}) Kerr black holes. 
We find an analytic expression for these black-hole resonances in terms of the black-hole 
{\it physical} parameters: its Bekenstein-Hawking temperature $T_{BH}$ 
and its horizon's angular velocity $\Omega$, 
which is valid in the intermediate asymptotic regime $1 \ll |\omega| \ll T_{BH}^{-1}$. 
\end{abstract}
\bigskip

]

Gravitational waves emitted by a perturbed black hole are dominated by
`quasinormal ringing', damped oscillations with a {\it discrete}
spectrum (see e.g., \cite{Nollert1} for a detailed review). 
At late times, all perturbations are
radiated away in a manner reminiscent of the last pure dying tones of
a ringing bell \cite{Press,Cruz,Vish,Davis}. 
Being the characteristic `sound' of
the black hole itself, these free oscillations 
are of great importance from 
the astrophysical point of view. They allow a direct way of identifying the spacetime 
parameters (especially, the mass and angular momentum of the black hole). 
This has motivated a flurry of activity with the aim of computing the spectrum of oscillations.

The ringing frequencies are located 
in the complex frequency plane characterized by $\mbox{Im}\,\omega <0$. It turns
out that for a Schwarzschild black hole, for a given angular harmonic index $l$ there exist an infinite number of quasinormal modes, for $n=0,1,2,\dots$, characterizing oscillations with decreasing
relaxation times (increasing imaginary part) \cite{Leaver,Bach}. On 
the other hand, the real part of the Schwarzschild black-hole resonances approaches an asymptotic 
{\it constant} value.

The QNMs have been the subject of much recent attention (see e.g., \cite{Kun,Motl,Cor,MotNei,CarLem,Hod2,KaRa,Abd,Ber1,Cho,Bri1,Gla,Nei,Mol,Pol,Bri2,LiHui,CarKon,BirCar,Swa,Pad,Bir,Ber2,Ber3,AndHow,Opp,Yos,MuSi,Rakn,LiZha,CaLeYo,MeMaVi,Tpad,DbiCar,RoPa,Mrs,CaLeYos,ZCGZ,JiJi,GoSu,VaZe,CaNaSc,DuWaSu,TaTa,CKi,OhSa,KeKuMe,EmVi,SoJi,LeSa,ShanDa,Wen,BrKoZh,NaSc,FerHol,ShSh,JinPan,Jing,GiMo,BerKok} 
and references therein), with the hope that these classical frequencies
may shed some light on the elusive theory of quantum gravity. 
These recent studies are motivated by an earlier work \cite{Hod1}, in which it was suggested to apply 
{\it Bohr's correspondence principle} to the black-hole resonances in order to determine the value of the 
fundamental area unit in a quantum theory of gravity. It was later suggested to use the black-hole 
QNM frequencies in order to fix the value of the Immirzi parameter in canonical quantum gravity \cite{Dreyer,Ash}. 
These proposals \cite{Hod1} have motivated a flurry of research attempting to calculate the {\it asymptotic} ringing 
frequencies of various types of black holes. 

Leaver \cite{Leaver} was the first to address the problem of computing the highly damped (asymptotic) ringing frequencies of the Schwarzschild black hole. Nollert \cite{Nollert2} found numerically (see also \cite{Andersson}) that the asymptotic (large $\mbox{Im}\,\omega$) behavior of the ringing frequencies of a Schwarzschild black hole is given by (we normalize $G=c=2M=1$)

\begin{equation}\label{Eq1}
\omega_n=0.0874247-{i \over 2} \left(n+{1 \over 2} \right)\  .
\end{equation}
Based on Bohr's correspondence principle, it was suggested \cite{Hod1} that 
this asymptotic (real) value actually equals $\ln(3)/(4\pi)$. 
This identity was further motivated by a heuristic picture, which was 
based on thermodynamical and statistical physics arguments \cite{Hod1,Muk,Beken1}. 
An analytical proof of this equality was later given in \cite{Motl}, 
This was followed by an analytical calculation of the asymptotic QNM frequencies of the 
(charged) Reissner-Nordstr\"om (RN) black hole \cite{MotNei}.

It should be emphasized however, that less is known about the corresponding 
QNM spectrum of the generic (rotating) Kerr black hole, which is the most interesting from a physical point of view. 
Former studies of the Kerr asymptotic spectrum \cite{Leaver,Det,Ono,Ber1,Ber2,Ber3} used numerical tools, 
and there are in fact no analytical results for the asymptotic Kerr QNM frequencies. 
In this work we provide analytical formulae for the intermediate asymptotic behavior of the Kerr QNM spectrum.
This is done by using a similarity between the intermediate QNMs of the Kerr black hole and the known asymptotic spectrum of the RN black hole. 

The dynamics of black-hole perturbations is governed by the Regge-Wheeler equation \cite{RegWheel} in the case of a Schwarzschild black hole, and by the Teukolsky equation \cite{Teukolsky} for the Kerr black hole. The black hole QNMs correspond to
solutions of the wave equations with the physical boundary
conditions of purely outgoing waves at spatial infinity 
and purely ingoing waves crossing the event horizon \cite{Detwe}. Such boundary 
conditions single out a {\it discrete} set of resonances $\{\omega_n\}$ (assuming a time dependence of the 
form $e^{-i\omega t}$). 
The solution to the radial Teukolsky equation may be expressed as \cite{Leaver} (assuming an azimuthal dependence of the from $e^{im\phi}$)

\begin{eqnarray}\label{Eq2}
R_{lm}& = &e^{i\omega r} (r-r_-)^{-1-s+i\omega+i\sigma_+} (r-r_+)^{-s-i\sigma_+}\nonumber \\
&&\Sigma_{n=0}^{\infty} d_n \Big({{r-r_+} \over {r-r_-}}\Big)^n\  ,
\end{eqnarray}
where $r_{\pm} =M \pm (M^2-a^2)^{1/2}$ are the black hole (event and inner) horizons, 
$a\equiv J/M$ is the black hole angular momentum per unit mass, 
and $\sigma_{+} \equiv (\omega r_{+}-ma)/(r_{+}-r_{-})$. 
The field spin-weight parameter $s$ takes the values $0, -1,$ and $-2$, respectively, 
for scalar, electromagnetic and gravitational fields. 

The sequence of expansion coefficients $\{d_n:n=1,2,\ldots\}$ is determined by a 
recurrence relation of the form \cite{Leaver}

\begin{equation}\label{Eq3}
\alpha_n d_{n+1}+\beta_n d_n +\gamma_n d_{n-1}=0\  ,
\end{equation}
with initial conditions $d_0=1$ and $\alpha_0 d_1+\beta_0 d_0=0$.
The recursion coefficients $\alpha_n,\beta_n$, and $\gamma_n$ are given in \cite{Leaver}. The quasinormal frequencies are determined by the requirement that the series in Eq. (\ref{Eq2}) is convergent, that is $\Sigma d_n$ exists and is finite \cite{Leaver}.

We find that the physical content of the recursion coefficients becomes clear when they are expressed in terms of the black-hole physical parameters: the Bekenstein-Hawking temperature $T_{BH}=(r_{+}-r_{-})/A$, 
and the horizon's angular velocity $\Omega=4\pi a/A$, where $A=4\pi(r_+^2+a^2)$ is the black-hole surface area. 
The recursion coefficients obtain a surprisingly simple form in terms of these physical quantities, 

\begin{equation}\label{Eq13}
\alpha_n=(n+1)(n+1-s-2i\beta_{+}\hat\omega)\  ,
\end{equation}

\begin{eqnarray}\label{Eq14}
\beta_n& = &-2(n+{1 \over 2}-2i\beta_{+}\hat\omega)
(n+{1 \over 2}-2i\omega r_{+})\nonumber \\
&&-(a\omega)^2+2ma\omega-s-{1 \over 2} -A_{lm}\  ,
\end{eqnarray}
and
\begin{equation}\label{Eq15}
\gamma_n=(n-2i\omega)(n+s-2i\beta_{+}\hat\omega)\  ,
\end{equation} 
where $\beta_{+} \equiv (4\pi T_{BH})^{-1}$ is the black-hole inverse temperature, 
$\hat\omega \equiv \omega-m\Omega$, and the angular separation constants $A_{lm}(a\omega)$ are given by an independent recurrence relation \cite{Leaver}.
To our best knowledge, this compact formulation of the recursion coefficients in terms 
of the black-hole {\it physical} parameters has not been done so far.

The Teukolsky equation also describes the propagation of scalar ($s=0$) waves in the RN spacetime [one should simply replace $r_{\pm}=M \pm (M^2-a^2)^{1/2}$ by 
$r_{\pm}=M \pm (M^2-Q^2)^{1/2}$, and take $a=0$ elsewhere]. 
The spectrum of the RN black hole may be found using a recursion relation of the form Eq. (\ref{Eq3}). The $\{\alpha_n\}$ and $\{\gamma_n\}$ coefficients of 
the RN perturbations have exactly the same form as in the Kerr case (where for the RN black hole $\hat\omega=\omega$). 
In addition, $\beta^{RN}_n=\beta^{Kerr}_n+(a\omega)^2-2ma\omega$ \cite{Note1}. 
In the intermediate asymptotic range, defined as $1 \ll |\omega| \ll T_{BH}^{-1}$, $\beta_n$ is dominated by its first term in Eq. (\ref{Eq14}). 
This term is at least of order $|\omega|/T_{BH}$, 
and is therefore much larger in magnitude than the $(a\omega)^2$ and $A_{lm}$ terms in this (intermediate asymptotic) regime. 
Hence, in this limit one finds that the $\{\beta_n\}$ coefficients of the Kerr black hole coincide with those of the RN black hole (in addition to the coincidence of the $\{\alpha_n\}$ and $\{\gamma_n\}$ terms). 

Thus, the intermediate asymptotic ($1 \ll |\omega| \ll T^{-1}_{BH}$) quasinormal frequencies of the near extremal Kerr black hole are related to the asymptotic frequencies of the RN black hole. 
The asymptotic spectrum of the RN black hole is determined, for $\mbox{Re}\,\omega>0$, by the equation \cite{MotNei,Note2}
\begin{equation}\label{Eq16}
2e^{-4\pi\beta_{+}\omega}+3e^{-4\pi\omega}=-1\ .
\end{equation}
The preceding discussion indicates that an expression similar to Eq. (\ref{Eq16}) must hold true for the intermediate asymptotic QNMs of the Kerr black hole. Equation (\ref{Eq16}) suggests that the spectrum depends on the combinations $\beta_+ \omega$ and $\omega$ appearing in Eqs. (\ref{Eq13})-(\ref{Eq15}), but does not depend explicitly on $\omega r_+$. Under this assumption, one finds that for a Kerr black hole \cite{Note3}

\begin{equation}\label{Eq17}
2e^{-4\pi\beta_{+}(\omega-m\Omega)}+3e^{-4\pi\omega}=-1\  .
\end{equation}
In the extremal limit $\beta_{+} \to \infty$, thus yielding 

\begin{equation}\label{Eq18}
\omega \to m\Omega+T_{BH}\ln2 -i 2\pi T_{BH} (n+{1\over2}) 
\end{equation}
for $m>0$, and 

\begin{equation}\label{Eq19}
\omega \to {{\ln3} \over {4\pi}} - {i\over 2} (n+{1 \over 2}) 
\end{equation}
for $m \le 0$. 
For $m=0$ there is an additional branch in cases where $\beta_+$ can be written as $\beta_+=2k/(2n+1)$, with $k$ and $n$ natural numbers. In this case, one finds 
\begin{equation}\label{Eq20}
\omega \to -{i \over 2}(n+{1 \over 2}) \ .
\end{equation}

These analytical results are consistent with previous numerical analyses \cite{Leaver,Ono}, which have began to probe the intermediate asymptotic regime. In particular, for $m>0$ the numerical results suggest that $\mbox{Re}\,\omega \to m \Omega$ in the extremal limit, and that in this limit the spacing between frequencies is $\Delta \omega \simeq i 2\pi T_{BH}$ \cite{Hod2}, in accord with the analytical result Eq. (\ref{Eq18}). For $m<0$, it was found numerically that $\Delta \omega \simeq i/2$ in the extremal limit, in agreement with Eq. (\ref{Eq19}). Two comments are in place. First, note that extrapolations of Eq. (\ref{Eq16}) from the RN to the Kerr black hole that differ from Eq. (\ref{Eq17}) will not agree with the numerical results. Second, the numerical results hold true for gravitational and electromagnetic perturbations, suggesting that Eqs. (\ref{Eq17})-(\ref{Eq20}) are valid for these perturbation fields as well. 

In summary, we have studied analytically the QNM spectrum of nearly extremal, 
rotating Kerr black holes. It was found that the intermediate asymptotic resonances 
can be expressed in terms of the black-hole {\it physical} parameters: 
its temperature $T_{BH}$, and its horizon's angular velocity $\Omega$, see Eqs. (\ref{Eq18})-(\ref{Eq20}). 
Our results may provide an important link towards an analytical calculation of the asymptotic spectrum, 
which is of great interest, both classically and quantum mechanically.

\bigskip
\noindent
{\bf ACKNOWLEDGMENTS}
\bigskip

The research of SH was supported by G.I.F. Foundation. We thank Emanuele Berti and 
Vitor Cardoso for helpful correspondence.


\begin{thebibliography}{99}

\bibitem{Nollert1} For an excellent review and a detailed list of references see 
H. P. Nollert, Class. Quantum Grav. {\bf 16}, R159 (1999).

\bibitem{Press} W. H. Press, Astrophys. J. {\bf 170}, L105 (1971).

\bibitem{Cruz} V. de la Cruz, J. E. Chase and W. Israel,
  Phys. Rev. Lett. {\bf 24}, 423 (1970).

\bibitem{Vish} C.V. Vishveshwara, Nature {\bf 227}, 936 (1970).

\bibitem{Davis} M. Davis, R. Ruffini, W. H. Press and R. H. Price,
  Phys. Rev. Lett. {\bf 27}, 1466 (1971).

\bibitem{Leaver} E. W. Leaver, Proc. R. Soc. A {\bf 402}, 285 (1985).

\bibitem{Bach} A. Bachelot and A. Motet-Bachelot,
  Ann. Inst. H. Poincar\'e {\bf 59}, 3 (1993).

\bibitem{Kun} G. Kunstatter, Phys. Rev. Lett. {\bf 90}, 161301 (2003).

\bibitem{Motl} L. Motl, Adv. Theor. Math. Phys. {\bf 6}, 1135 (2003).

\bibitem{Cor} A. Corichi, Phys. Rev. D {\bf 67}, 087502 (2003).

\bibitem{MotNei} L. Motl and A. Neitzke, Adv. Theor. Math. Phys. {\bf 7}, 307 (2003).

\bibitem{CarLem} V. Cardoso and  J. P. S. Lemos, Phys. Rev. D {\bf 67}, 084020 (2003).

\bibitem{Hod2} S. Hod, Phys. Rev. D {\bf 67}, 081501 (2003).

\bibitem{KaRa} R. K. Kaul and S. K. Rama, Phys. Rev. D {\bf 68}, 024001 (2003).

\bibitem{Abd} E. Abdalla, K. H. C. Castello-Branco, and A. Lima-Santos, Mod. Phys. Lett. A {\bf 18}, 1435 (2003).

\bibitem{Ber1} E. Berti and K. D. Kokkotas, Phys. Rev. D {\bf 68}, 044027 (2003).

\bibitem{Cho} H. T. Cho, Phys. Rev. D {\bf 68}, 024003 (2003).

\bibitem{Bri1} A. M. van den Brink, J. Math. Phys. {\bf 45}, 327 (2004).

\bibitem{Gla} K. Glampedakis and N. Andersson, Class. Quant. Grav {\bf 20}, 3441 (2003).

\bibitem{Nei} A. Neitzke, e-print hep-th/0304080. 

\bibitem{Mol} C. Molina, Phys. Rev D {\bf 68}, 064007 (2003).

\bibitem{Pol} A. P. Polychronakos, Phys. Rev. D {\bf 69}, 044010 (2004).

\bibitem{Bri2} A. M. van den Brink, Phys. Rev. D {\bf 68}, 047501 (2003).

\bibitem{LiHui} L. H. Xue, Z. X. Shen, B. Wang, and R. K. Su, Mod. Phys. Lett. A {\bf 19}, 239 (2004).

\bibitem{CarKon} V. Cardoso, R. Konoplya, J. P. S. Lemos, Phys. Rev. D {\bf 68}, 044024 (2003).

\bibitem{BirCar} D. Birmingham, S. Carlip, and Y. Chen, Class. Quant. Grav. {\bf 20}, L239 (2003).

\bibitem{Swa} J. Swain, Int. J. Mod. Phys. D {\bf 12}, 1729 (2003).

\bibitem{Pad} T. Padmanabham and A. Patel, e-print hep-th/0305165.

\bibitem{Bir} D. Birmingham, Phys. Lett. B {\bf 569}, 199 (2003).

\bibitem{Ber2} E. Berti, V. Cardoso, K. D. Kokkotas and H. Onozawa, Phys. Rev. D {\bf 68}, 124018 (2003).

\bibitem{Ber3} E. Berti, V. Cardoso, S. Yoshida, Phys. Rev. D {\bf 69}, 124018 (2004).

\bibitem{AndHow} N. Andersson and C. J. Howls, Class. Quant. Grav. {\bf 21}, 1623 (2004).

\bibitem{Opp} J. Oppenheim, Phys. Rev. D {\bf 69}, 044012 (2004).

\bibitem{Yos} S. Yoshida and T. Futamase, Phys. Rev. D {\bf 69}, 064025 (2004).

\bibitem{MuSi} S. Musiri and G. Siopsis, Class. Quant. Grav. {\bf 20}, L285 (2003); 
Phys. Lett. B {\bf 579}, 25 (2004).

\bibitem{Rakn} R. A. Konoplya, Phys. Rev. D {\bf 68}, 124017 (2003); 
Phys. Rev. D {\bf 70}, 047503 (2004); Phys. Rev. D {\bf 71}, 024038 (2005).

\bibitem{LiZha} Y. Ling and H. Zhang, Phys. Rev. D {\bf 68}, 101501 (2003).

\bibitem{CaLeYo} V. Cardoso, J. P. S. Lemos, and S. Yoshida, Phys. Rev. D {\bf 69}, 044004 (2004).

\bibitem{MeMaVi} A. J. M. Medvev, D. Martin, and M. Visser, 
Class. Quant. Grav. {\bf 21}, 1393 (2004); Class. Quant. Grav. {\bf 21}, 2393 (2004).

\bibitem{Tpad} T. Padmanabham, Class. Quant. Grav. {\bf 21}, L1 (2004).

\bibitem{DbiCar} D. Birmingham and S. Carlip, Phys. Rev. Lett {\bf 92}, 111302 (2004).

\bibitem{RoPa} T. R. Choudhury and T. Padmanabham, Phys. Rev. D {\bf 69}, 064033 (2004).

\bibitem{Mrs} M. R. Setare, Class. Quant. Grav. {\bf 21}, 
1453 (2004); Phys. Rev. D {\bf 69}, 044016 (2004).

\bibitem{CaLeYos} V. Cardoso, J. P. S. Lemos, and S. Yoshida, 
JHEP {\bf 312}, 41 (2003).

\bibitem{ZCGZ}, H. Zhang, Z. Cao, X. Gong, and W. Zhou, Class. Quant. Grav. {\bf 21}, 917 (2004).

\bibitem{JiJi} J. Jing, Phys. Rev. D {\bf 69}, 084009 (2004).

\bibitem{GoSu} G. Gour and V. Suneeta, Class. Quant. Grav. {\bf 21}, 3405 (2004).

\bibitem{VaZe} L. Vanzo and S. Zerbini, Phys. Rev. D {\bf 70}, 044030 (2004).

\bibitem{CaNaSc} V. Cardoso, J. Natario, and R. Schiappa, J. Math. Phys. {\bf 45}, 4698 (2004).

\bibitem{DuWaSu} D. P. Du, B. Wang, and R. K. Su, Phys. Rev. D {\bf 70}, 064024 (2004).

\bibitem{TaTa} T. Tamaki and H. Nomura, Phys. Rev. D {\bf 70}, 044041 (2004).

\bibitem{CKi} C. Kiefer, Class. Quant. Grav. {\bf 21}, L123 (2004).

\bibitem{OhSa} A. Ohashi and M. Sakagami, Class. Quant. Grav. {\bf 21}, 3973 (2004).

\bibitem{KeKuMe} J. Kettner, G. Kunstatter, and A. J. M. Medved, Class. Quant. Grav. {\bf 21}, 5317 (2004).

\bibitem{EmVi} E. Berti, V. Cardoso, and J. P. S. Lemos, Phys. Rev. D {\bf 70}, 124006 (2004).

\bibitem{SoJi} S. Chen and J. Jing, Class. Quant. Grav. {\bf 22}, 533 (2005).

\bibitem{LeSa} S. Lepe and J. Saavedra, e-print gr-qc/0410074.

\bibitem{ShanDa} S. Das and S. Shankaranarayanan, Class. Quant. Grav. {\bf 22}, L7 (2005).

\bibitem{Wen} F. W. Shu and Y. G. Shen, Phys. Rev. D {\bf 70}, 084046 (2004).

\bibitem{BrKoZh} K. H. C. Castello-Branco, R. A. Konoplya, and 
A. Zhidenko, Phys. Rev. D {\bf 71}, 047502 (2005).

\bibitem{NaSc} J. natario and R. Schiappa, e-print hep-th/0411267.

\bibitem{FerHol} S. Fernando and C. Holbrook, e-print hep-th/0501138.

\bibitem{ShSh} F. W. Shu and Y. G. Shen, e-print gr-qc/0501098.

\bibitem{JinPan} J. Jing and Q. Pan, e-print gr-qc/0502011.

\bibitem{Jing} J. Jing, e-print gr-qc/0502023; e-print gr-qc/0502010.

\bibitem{GiMo} M. Giammatteo and I. G. Moss, e-print gr-qc/0502046.

\bibitem{BerKok} E. Berti and K. D. Kokkotas, e-print gr-qc/0502065. 

\bibitem{Hod1} S. Hod, Phys. Rev. Lett. {\bf 81}, 4293 (1998).

\bibitem{Dreyer} O. Dreyer, Phys. Rev. Lett. {\bf 90}, 081301 (2003).

\bibitem{Ash} A. Ashtekar, J. C. Baez, A. Corichi and K. Krasnov, 
Phys. Rev. Lett. {\bf 80}, 904 (1998).

\bibitem{Nollert2} H. P. Nollert, Phys. Rev. D {\bf 47}, 5253 (1993).

\bibitem{Andersson} N. Andersson, Class. Quantum Grav. {\bf 10}, L61 (1993).

\bibitem{Muk} V. Mukhanov, JETP Lett. {\bf 44}, 63 (1986).

\bibitem{Beken1} J. D. Bekenstein, Phys. Rev. D {\bf 7}, 2333 (1973).

\bibitem{Det} S. Detweiler, Astrophys. J. {\bf 239}, 292 (1980).

\bibitem{Ono} H. Onozawa, Phys. Rev. D {\bf 55}, 3593 (1997).

\bibitem{RegWheel} T. Regge and J. A. Wheeler, Phys. Rev. {\bf 108},
  1063 (1957).

\bibitem{Teukolsky} S. A. Teukolsky, Phys. Rev. Lett. {\bf 29}, 1114 (1972); 
Astrophys. J. {\bf 185}, 635 (1973).

\bibitem{Detwe} S. L. Detweiler, in Sources of Gravitational
  Radiation, edited by L. Smarr (Cambridge University Press,
  Cambridge, England, 1979).

\bibitem{Note1} 
For $|a \omega|\gg 1$, the angular wave functions that satisfy the angular Teukolsky equation approach the prolate spheroidal wave functions, implying that $A_{lm} = O(\alpha \omega)$ \cite{Abr,Flam,Ber3}. 
Thus, $A_{lm}$ is subdominent to the $(a\omega)^2$ term in $\beta_n^{Kerr}$. 

\bibitem{Abr} M. Abramowitz, {\it Handbook of Mathematical Functions}, (Dover 
Publications, inc., New York, 1965).

\bibitem{Flam} C. Flammer, in {\it Spheroidal Wave Functions}, (Stanford University Press, Stanford, CA, 1957).

\bibitem{Note2} The final result of \cite{MotNei} for the RN spectrum, 
can be written as Eq. (\ref{Eq16}) by simply multiplying both sides of 
equation $(66)$ in \cite{MotNei} by $e^{-\beta_{+} \omega}$.

\bibitem{Note3} This relation is valid for $\mbox{Re}\,\omega>0$. The frequencies with $\mbox{Re}\,\omega<0$ may be determined from the well-known symmetry of the Kerr spacetime \cite{Leaver}: if $\omega_{l,m}$ is a QNM frequency, then $-\omega^{*}_{l,-m}$ is 
also a solution.

\end{thebibliography}
\end{document}